\documentclass[12pt]{article}
\usepackage{graphicx}
\usepackage{setspace}
\usepackage[margin=1in]{geometry}
\usepackage{subcaption}
\usepackage{bm} 
\usepackage{booktabs}
\usepackage{amsmath, amsfonts, amssymb}
\usepackage{threeparttable} 
\usepackage{enumerate}
\usepackage{float}
\usepackage{xurl}
\usepackage[
            hyperref=true, 
            doi=false,
            url=false, 
            isbn=false, 
            eprint=false,
            style=authoryear,
            maxbibnames=99,
            maxcitenames=99,
            ]{biblatex}
\renewbibmacro{in:}{}
\usepackage[
  hidelinks
]{hyperref}
\usepackage[all]{hypcap}
\hypersetup{pageanchor=false}

\title{Tracking the Economy through Firm Creation: Evidence from Real-Time Administrative Data\footnotetext[0]{\textbf{Acknowledgements:} We thank the following people for comments: Daniel Albuquerque, Maria Balgova, Gökçe Karasoy Can, Aubrey Poon, Maren Froemel, Sophie Piton, Ida Hjortso, Jonathan Haskel, Andrea Alati, Alberto Polo, Ambrogio Cesa-Bianchi, Ben Hemingway, Rebecca Riley, Javier Miranda, Neeltje Van Horen, Misa Tanaka and Petr Sedláček. We thank conference and seminar participants at: Bristol, Bank of England, ESCoE Manchester 2024, CAED Penn State 2024, UNSW-ESCoE 2024, NIESR, University of Westminster and The Productivity Institute 2025. 
}
}
\author{Yannis Galanakis\thanks{King's College London, ESCoE \& TPI; yannis.galanakis@kcl.ac.uk} \and Anthony Savagar\thanks{Bank of England, University of Kent \& ESCoE; asavagar@gmail.com. The views expressed in this paper are those of the authors, and not necessarily those of the Bank of England or its committees.}
}

\date{\today}

\begin{document}
\begin{titlepage}
\maketitle

\begin{abstract}
We introduce a novel real-time dataset—\emph{Companies House Real-Time} (CHRT)—that captures daily firm creation and dissolution activity for the full population of UK-registered companies. CHRT provides a timely measure of business formation, becoming available months before official business demography statistics. We show that incorporation activity leads taxable business births and contains forward-looking information about employment and output growth. Consistent with this, a structural vector autoregression (SVAR) indicates that positive shocks to firm entry generate persistent increases in employment and output.
\end{abstract}
\end{titlepage}

This paper shows that real-time firm incorporation data provide a timely and informative indicator of future macroeconomic conditions. We construct a novel dataset based on UK Companies House registrations and show that firm entry leads taxable business creation statistics and predicts future employment and output growth.

Timely measures of business formation are important for monitoring macroeconomic conditions. However, official business creation statistics are typically released with publication lags, as they aim to capture firms once they become economically active rather than just legal registrations. In the UK, official Business Demography (BD) statistics are based on VAT and PAYE registrations and therefore record firms only when they become liable for these taxes. This measurement approach overlooks legal registrations that may provide an early signal of future economic conditions.

This paper addresses that limitation using a new source of high-frequency administrative data. In the UK, all incorporated firms are required to register with Companies House, the official registrar of companies. Over the last decade, the registration process has become almost entirely digital, allowing near-instantaneous recording of firm incorporations. Companies House data are released publicly at daily frequency and provide broad coverage across sectors and regions. 

We construct the \emph{Companies House Real-Time} (CHRT) dataset, a new administrative dataset measuring firm incorporation activity in real time. CHRT combines Companies House register snapshots with API records to provide daily information on firm incorporations by industry and geography since 2012. The dataset allows high-frequency monitoring of business formation using legally defined incorporation events. We provide a dashboard displaying CHRT indicators by sector and region:
\newline
\url{https://asavagar.shinyapps.io/BusinessDynamicsDashboard/}.

Firm incorporation activity may lead economic conditions for two complementary reasons. First, incorporation is the earliest observable stage in the creation of future firms and therefore may provide an early signal of subsequent business births, employment, and output. Second, incorporation decisions may reveal entrepreneurs' expectations about future economic conditions. In standard models of firm entry, entrepreneurs enter when the expected discounted value of future profits exceeds entry costs. Consequently, fluctuations in incorporation activity may provide an early signal of future economic outcomes, either because newly created firms later become economically active or because entrepreneurs respond to future economic conditions before they appear in conventional macroeconomic data.

We make three contributions. First, we construct and validate a novel real-time dataset on UK firm incorporation activity using Companies House administrative records, which we make publicly available. Second, we show that incorporation activity leads official ONS business demography (tax-based births) statistics by several quarters and contains forward-looking information about employment and output growth. Third, we show that incorporation growth improves employment forecasts and that shocks to firm entry have persistent effects on employment and output. Quantitatively, a one-standard-deviation increase in year-on-year firm entry growth (14.5\%) raises year-on-year employment growth by 0.4\% percentage points one year later.

Methodologically, we use parsimonious estimation techniques and transparent detrending to highlight the raw information content of the data. We implement lead-lag regressions, autoregressive forecasts, and Cholesky VARs on monthly variables (entry, employment, output), comparable to exercises in \textcite{GourioMesserSiemer2016_AER, AsturiasDinlersozHaltiwangerHutchinson2023_AEApp}. We work with month-on-month and year-on-year differenced data to distinguish short-run fluctuations in firm entry from more persistent changes associated with entrepreneurial cycles.

\subsubsection*{Related Literature}

This paper contributes to two strands of research: the macroeconomic effects of firm entry and the use of real-time business formation data to measure economic activity.

\paragraph{Macroeconomic effects of firm entry}

A large literature studies the role of firm entry in business cycles, labour market dynamics, and long-run growth \parencite{BilbiieGhironiMelitz2012_JPE, LewisPoilly2012_JME, LewisStevens2015_EER, HamanoZanetti2017_RED, LewisWinkler2017_IER, SedlacekSterk2017_AER, acemoglu2018innovation, Sedlacek2020_JME, SterkSedlacekPugsley2021_AER, GutierrezJonesPhilippon2021_JME, HamanoZanetti2022_EER}. Our work is most closely related to \textcite{GourioMesserSiemer2016_AER}, who estimate the effect of firm entry on US GDP using annual state-level data. We contribute by showing that high-frequency, real-time incorporation data contain forward-looking information about employment and output growth. During the COVID-19 period, related work used business formation measures to study economic resilience and recovery across countries \parencite{Benedetti-FasilSedlacekSterk2022_EP, DeckerHaltiwanger2023_BPEA, BahajPitonSavagar2024_EP}.

\paragraph{Real-time business formation data}

Our paper also contributes to a growing literature using administrative data to monitor business formation in real time. In the United States, \textcite{BayardDinlersozDunneHaltiwangerMirandaStevens2018_NBER} introduce the Business Formation Statistics (BFS), based on Employer Identification Number (EIN) applications. Recent work uses BFS data to study employer transitions, business dynamism, and macroeconomic activity \parencite{DinlersozDunneHaltiwangerPenciakova2021_AEAPP, DeckerHaltiwanger2023_Brookings}. Most closely related to our paper, \textcite{AsturiasDinlersozHaltiwangerHutchinson2023_AEApp} show that real-time business registrations in the United States contain forward-looking information about macroeconomic activity. We provide analogous evidence for the UK using a novel real-time incorporation dataset. Unlike BFS, which measures prospective business formation through tax applications, Companies House records the legal act of incorporation directly and provides daily information on firm creation together with detailed industry and geographic classifications.

\subsubsection*{Roadmap}

Section \ref{sec:data} describes the CHRT dataset. Section \ref{sec:leading_indicator} presents lead--lag and forecasting evidence. Section \ref{sec:var} analyses the dynamic relationship between firm entry and macroeconomic activity. Section \ref{sec:conclusion} concludes.

\section{Data and Measurement}
\label{sec:data}

\subsection{The Companies House Real-Time (CHRT) Dataset}

We construct a novel high-frequency dataset of UK firm creation using publicly available records from Companies House. The Companies House Real-Time (CHRT) dataset combines historical register snapshots with information from the Companies House API to provide near real-time coverage of incorporated firms since 2012. The data contain firm identifiers, incorporation and dissolution dates, five-digit SIC codes, and postcode-level geographic information.

Active firms are defined using the Companies House register of live companies. Entrants are identified from new incorporations appearing between successive register snapshots, while exits are dated using official dissolution records from the Companies House API. However, the leading feature of incorporations does not transfer to dissolutions. Addition and removal from the Companies House register bookends the firm lifecyle process. That is to setupa company, the first step is to register:
\[
\text{CH Incorporation}
\rightarrow
\text{Economic Activity}
\rightarrow
\text{VAT/PAYE Registration}
\]
Whereas to remove a company from the register, the final step is to de-register (dissolve):
\[
\text{Economic Activity Ceases}
\rightarrow
\text{VAT/PAYE Inactivity}
\rightarrow
\text{CH Dissolution}
\]

The CHRT data offer several advantages relative to official business statistics. First, they are highly timely, allowing firm creation to be observed close to the point of legal registration. Second, they provide near-universal coverage of incorporated firms with detailed sectoral and geographic information. However, Companies House records legal incorporation rather than economic activity per se, implying that some live companies may be economically dormant. In addition, the data exclude unincorporated businesses and self-employed individuals. 

Appendix~\ref{sec:appendix_chrt} provides additional details on data construction, validation, and institutional features of the Companies House register.

\subsection{Official Business Birth Statistics}

We compare CHRT against quarterly business demography statistics from the Office for National Statistics (ONS). These data are based on the Inter-Departmental Business Register (IDBR), which includes businesses registered for VAT or PAYE and therefore captures economically active firms rather than all legal incorporations.

The IDBR differs conceptually from Companies House in two key respects. First, it includes a broader range of legal forms, including unincorporated businesses and self-employed firms. Second, firms appear only after completing tax registration with HM Revenue and Customs (HMRC), creating an administrative lag between legal incorporation and inclusion in official statistics. According to \textcite{ONS2022RegistrationLags}, around 85\% of firms that eventually appear on the IDBR do so within twelve months of incorporation. CHRT therefore provides a substantially more timely measure of firm formation dynamics.

\subsection{Macroeconomic Data}

We use monthly macroeconomic data from the ONS to study the relationship between firm creation and economic activity. Our employment measure is the monthly PAYE Real Time Information (RTI) series from HMRC and the ONS, which provides seasonally adjusted counts of payrolled employees from July 2014 onwards. The data are constructed from administrative payroll records submitted by employers each time employees are paid, providing near-universe coverage of employees with minimal publication lag.\footnote{Employment data: Table 1 from \url{https://www.ons.gov.uk/employmentandlabourmarket/peopleinwork/earningsandworkinghours/bulletins/earningsandemploymentfrompayasyouearnrealtimeinformationuk/june2025}, retrieved June 10, 2025.}

As our measure of aggregate economic activity, we use the ONS monthly Gross Value Added (GVA) series, specifically the ``Gross Value Added -- Monthly (Index 1dp): CVM SA'' measure. This chained-volume, seasonally adjusted index provides a timely indicator of short-run economic activity between quarterly national accounts releases. Monthly GVA estimates are, however, more volatile and subject to revision than official quarterly national accounts.\footnote{Output data: Columns 1 and 2 from \url{https://www.ons.gov.uk/economy/grossdomesticproductgdp/datasets/gdpmonthlyestimateuktimeseriesdataset}.}

\subsection{Variable Construction and Descriptive Statistics}

We consider both month-on-month and year-on-year growth rates of firm entry, employment, and output. Month-on-month growth rates are defined as $\Delta \log X_t = \log X_t - \log X_{t-1}$ and, later, are typically estimated in models with month fixed effects to control for seasonality. Year-on-year growth rates are defined as $\Delta_{12}\log X_t = \log X_t - \log X_{t-12}$ and instead compare each monthly observation to the same month one year earlier, thereby differencing out stable seasonal patterns directly.

Although the same transformations are applied across variables, their economic interpretation differs somewhat. Firm entry is a monthly flow variable, while employment and output are monthly levels. More generally, month-on-month growth focuses on short-run fluctuations, whereas year-on-year growth abstracts from some high-frequency volatility and therefore captures relatively slower-moving changes in macroeconomic conditions.

\begin{table}[H]
\centering
\caption{Summary statistics}
\label{tab:desc_stats_main}
\begin{tabular}{lccc}
\toprule
Variable & Mean & Std. Dev. & N \\
\midrule
$E_t$                     & 62,094 & 10,894 & 140 \\
$L_t$                     & 29.0m  & 1.1m   & 140 \\
$Y_t$                     & 95.6   & 6.5    & 140 \\
\midrule
$100 \times \Delta \log E_t$      & 0.1 & 15.0 & 139 \\
$100 \times \Delta \log L_t$      & 0.1 & 0.7  & 139 \\
$100 \times \Delta \log Y_t$      & 0.1 & 3.0  & 139 \\
\midrule
$100 \times \Delta_{12}\log E_t$  & 3.0 & 14.5 & 128 \\
$100 \times \Delta_{12}\log L_t$  & 1.0 & 2.3  & 128 \\
$100 \times \Delta_{12}\log Y_t$  & 2.1 & 5.8  & 128 \\
\bottomrule
\end{tabular}

\vspace{0.3em}
\footnotesize
\textit{Notes:} $E_t$ denotes monthly firm incorporations, $L_t$ PAYE employment, and $Y_t$ monthly GVA. Growth rates are log differences multiplied by 100 and are therefore approximately percentage changes. Full distributional statistics are reported in Appendix Table~\ref{tab:incl_covid_stats}.
\end{table}

Table \ref{tab:desc_stats_main} reports descriptive statistics for the levels and growth rates of firm entry, employment, and output. Entry growth is substantially more volatile than either employment or output growth. The standard deviation of month-on-month entry growth is 15\%, compared with 0.7\% for employment and 3\% for output. A similar pattern is evident using year-on-year growth rates. While year-on-year transformations smooth all three series by removing seasonal variation and some high-frequency noise, entry remains considerably more volatile than the corresponding employment and output measures. 

Additional descriptive evidence, including time-series plots and descriptive statistics excluding the COVID-19 period, is reported in Appendix \ref{app:descriptive}.

\section{Firm Entry as a Leading Indicator}
\label{sec:leading_indicator}

This section evaluates whether real-time incorporation data contain forward-looking information about business formation and macroeconomic activity. We first compare CHRT incorporations with official ONS business demography statistics and then examine whether firm entry predicts subsequent employment and output growth using lead--lag regressions and out-of-sample forecasting exercises.

\subsection{From Incorporations to Business Births}
\label{subsec:chrt_vs_bd}

We first assess whether real-time incorporation data contain information about subsequent official measures of business formation. As discussed in Section \ref{sec:data}, the CHRT dataset records firms at the point of legal incorporation, whereas official business demography statistics only include firms once they meet VAT or PAYE registration thresholds. This institutional difference implies that Companies House registrations should lead official business birth measures.

\begin{figure}[H]
\centering
\includegraphics[width=0.82\textwidth]{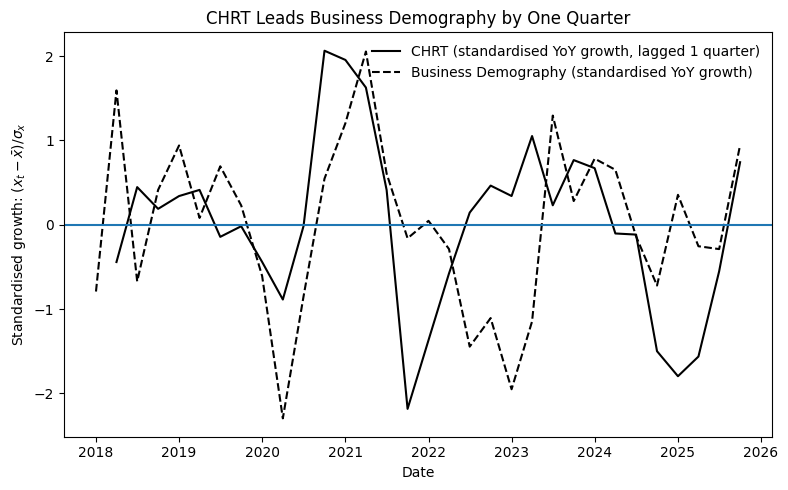}
\caption{
CHRT as a leading indicator of Business Demography births. Both series are year-on-year log changes, standardised to have mean zero and unit variance. CHRT incorporations are lagged by one quarter to illustrate their relationship with subsequent official business births.
}
\label{fig:chrt_bd_standardised}
\end{figure}

Figure \ref{fig:chrt_bd_standardised} plots standardised year-on-year growth in BD births together with standardised year-on-year growth in CHRT incorporations lagged by one quarter. The two series display strong co-movement once CHRT is shifted forward, consistent with incorporation data providing an early signal of subsequent official business births.

To quantify this relationship, we estimate lead--lag regressions between CHRT incorporations and ONS Business Demography (BD) births. Since BD is observed at quarterly frequency, we work with year-on-year log changes,
\[
\Delta_{4}\log x_t = \log x_t - \log x_{t-4},
\]
which remove seasonality and focus on medium-run changes in business formation. Specifically, we estimate
\[
\Delta_{4}\log BD_{t+h}
=
\alpha_h
+
\beta_h \Delta_{4}\log CHRT_t
+
\varepsilon_{t+h},
\]
for horizons $h=0,1,\ldots,6$ quarters. A positive coefficient for $h>0$ implies that CHRT contains information about future official business births.

\begin{figure}[H]
\centering
\includegraphics[width=0.82\textwidth]{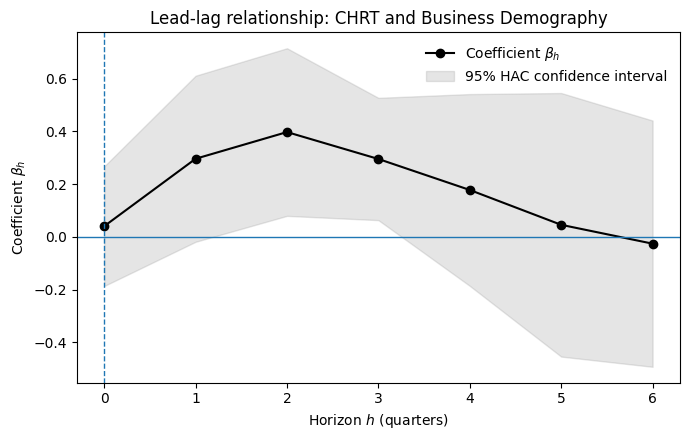}
\caption{
Lead--lag relationship between CHRT incorporations and ONS Business Demography births. The figure reports coefficients from regressions of year-on-year log changes in BD births at horizon $t+h$ on year-on-year log changes in CHRT incorporations at time $t$. Shaded areas denote 95\% HAC confidence intervals.
}
\label{fig:chrt_bd_leadlag}
\end{figure}

Figure \ref{fig:chrt_bd_leadlag} reports the estimated coefficients together with 95\% HAC confidence intervals. The relationship strengthens at short positive horizons, with coefficients rising from close to zero contemporaneously to around 0.3--0.4 after two to three quarters before gradually declining. This pattern is consistent with the administrative structure of the data: incorporations are recorded immediately in CHRT, while firms only enter Business Demography statistics once they become sufficiently economically active to register for VAT or PAYE. Quantitatively, the estimates imply that a 1\% increase in the annual growth rate of CHRT incorporations is associated with roughly a 0.4\% increase in the annual growth rate of official business births around two quarters later.

Overall, the results suggest that incorporation activity contains forward-looking information about subsequent business formation. The year-on-year specification therefore captures the medium-run transition from legal incorporations to economically active firms recorded within the VAT/PAYE-based business population.

\subsection{Predicting Employment and Output}
\label{subsec:lead_lag}

We next assess whether firm entry contains forward-looking information about aggregate economic activity. Incorporation decisions depend on expectations of future profitability, demand, and financing conditions, and may therefore adjust before realised macroeconomic outcomes.

\begin{figure}[H]
    \centering
    \includegraphics[width=\textwidth]{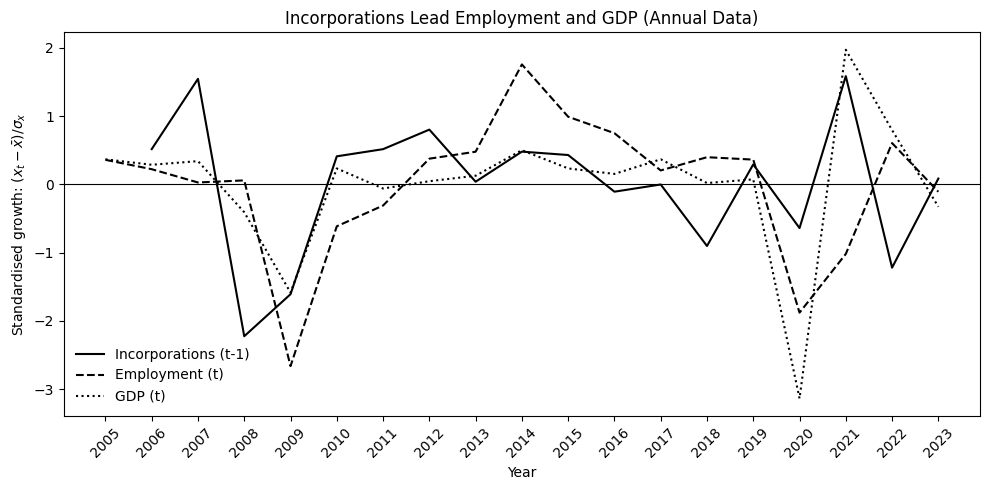}
    \caption{Firm creation as a leading indicator of output and employment growth}
    \label{fig:firmcreation_growth}
    \caption*{\scriptsize \textbf{Notes:} The figure plots standardised annual growth rates of firm incorporations, GVA, and total employment in the UK over 2005--2023. Each series is transformed as $(x_t - \bar{x})/\sigma_x$ to facilitate comparison of cyclical fluctuations. Incorporations are based on Companies House annual statistics (financial year basis). GDP growth is year-on-year real GDP growth (CVM, seasonally adjusted) from the ONS. Employment growth is the annual percentage change in total employment (aged 16+, seasonally adjusted).}
\end{figure}

Figure \ref{fig:firmcreation_growth} plots standardised annual growth rates of firm incorporations alongside output and employment growth in the UK since 2005.\footnote{Each series is standardised as $(x_t - \bar{x})/\sigma_x$, so that it has zero mean and unit variance. Values therefore represent deviations from the historical average in units of standard deviations.} The figure suggests that movements in incorporations often precede turning points in output and employment growth. In particular, declines in incorporations occur ahead of the contractions associated with the Global Financial Crisis and the COVID-19 shock, while recoveries in entry are followed by subsequent improvements in macroeconomic conditions.

We estimate two lead--lag specifications relating future employment and output growth to contemporaneous firm entry growth. The first uses year-on-year log growth rates:

\begin{equation}
\Delta_{12}\log Z_{t+h}
=
\alpha_h
+
\beta_h \Delta_{12}\log E_t
+
\varepsilon_{t+h},
\end{equation}

where $Z_t \in \{L_t,Y_t\}$ denotes employment or output, and $\Delta_{12}\log X_t = \log X_t - \log X_{t-12}$. The second specification uses monthly log differences with month fixed effects:

\begin{equation}
\Delta \log Z_{t+h}
=
\alpha_h
+
\beta_h \Delta \log E_t
+
\gamma_m
+
\varepsilon_{t+h},
\end{equation}

where $\Delta \log X_t = \log X_t - \log X_{t-1}$ and $\gamma_m$ denotes month fixed effects. In both cases, horizons $h$ are measured in months ahead and HAC/Newey--West standard errors are used for inference.

\begin{table}[H]
\centering
\caption{Lead--lag relationship between firm entry and macroeconomic outcomes}
\label{tab:leadlag_results}
\begin{tabular}{lcccccc}
\toprule
& \multicolumn{3}{c}{Employment Growth ($L$)}
& \multicolumn{3}{c}{Output Growth ($Y$)} \\
\cmidrule(lr){2-4} \cmidrule(lr){5-7}
Horizon ($h$) & Coef. & $p$-value & Obs.
              & Coef. & $p$-value & Obs. \\
\midrule
\multicolumn{7}{l}{\textit{Annual log differences: $\Delta_{12}\log$}} \\
0  & -0.031 & 0.185 & 128 & -0.043 & 0.281 & 128 \\
6  & -0.006 & 0.702 & 122 &  0.039 & 0.428 & 122 \\
10 &  0.027$^{*}$   & 0.050 & 118 &  0.160$^{*}$   & 0.082 & 118 \\
11 &  0.033$^{**}$  & 0.034 & 117 &  0.126$^{*}$   & 0.060 & 117 \\
12 &  0.034$^{**}$  & 0.028 & 116 &  0.046 & 0.270 & 116 \\
\midrule
\multicolumn{7}{l}{\textit{Monthly log differences with month fixed effects: $\Delta \log$}} \\
0  & -0.001 & 0.444 & 139 &  0.063 & 0.128 & 139 \\
6  &  0.000 & 0.982 & 133 & -0.016$^{*}$ & 0.085 & 133 \\
10 &  0.002 & 0.160 & 129 &  0.035 & 0.167 & 129 \\
11 &  0.002 & 0.526 & 128 &  0.007 & 0.643 & 128 \\
12 &  0.003$^{**}$ & 0.029 & 127 & -0.008 & 0.529 & 127 \\
\bottomrule
\end{tabular}

\vspace{0.5em}

\begin{minipage}{0.92\textwidth}
\footnotesize
\textit{Notes:} The table reports coefficients from separate lead--lag regressions of future employment and output growth on current firm entry growth at horizon $h$. The upper panel uses annual log differences, while the lower panel uses monthly log differences with month fixed effects. HAC robust standard errors are used. $^{*}$, $^{**}$, and $^{***}$ denote significance at the 10\%, 5\%, and 1\% levels respectively.
\end{minipage}
\end{table}

Table \ref{tab:leadlag_results} shows that firm entry growth has little contemporaneous relationship with employment growth but becomes positively associated with employment growth roughly one year later. In the year-on-year specification, a 10\% increase in entry growth is associated with approximately 0.3 percentage points higher employment growth twelve months later. Given that the standard deviation of year-on-year entry growth is approximately 14.5 percentage points (Table \ref{tab:desc_stats_main}), a one-standard-deviation increase in entry growth predicts roughly 0.4--0.5 percentage points higher employment growth after one year. These dynamics are consistent with firms first incorporating and only gradually expanding hiring activity.

The relationship between entry growth and output growth is weaker and less precisely estimated, although positive coefficients emerge at horizons around 10--11 months. More generally, the year-on-year specification produces stronger and more persistent relationships than the month-on-month specification.

The two specifications capture different frequencies of variation. Year-on-year growth rates remove seasonality and smooth high-frequency volatility, making slower-moving relationships easier to detect. Month-on-month growth rates instead emphasise short-run fluctuations, which are likely to contain more transitory variation and timing noise.

We complement the lead--lag analysis with an out-of-sample forecasting exercise that assesses whether incorporation growth contains predictive information beyond the persistence already embedded in macroeconomic aggregates. For each horizon $h \in \{3,6,12\}$ months, we estimate direct forecasting regressions of the form

\begin{equation}
\Delta_k \log Z_{t+h}
=
\alpha_h
+
\beta_h \Delta_k \log Z_t
+
\gamma_h \Delta_k \log E_t
+
\varepsilon_{t+h},
\end{equation}

where $Z_t \in \{L_t,Y_t\}$ denotes employment or output, $E_t$ denotes incorporations, and $\Delta_k$ is either a 12-month log difference ($k=12$) or a monthly log difference ($k=1$). Forecast performance is evaluated relative to an autoregressive benchmark using out-of-sample root mean squared error (RMSE).

\begin{table}[H]
\centering
\caption{Forecasting performance of incorporation growth}
\label{tab:forecast_results}
\begin{threeparttable}
\footnotesize
\setlength{\tabcolsep}{4pt}
\begin{tabular}{llccc}
\toprule
Outcome & $h$ & RMSE gain (\%) & Entry coef. & $p$-value \\
\midrule

\multicolumn{5}{l}{\textit{Panel A: 12-month growth rates}} \\

Output         & 3  & -50.8  & 0.0223  & 0.724 \\
Output         & 6  & -168.8 & 0.0695  & 0.395 \\
Output         & 12 & -12.8  & 0.0335  & 0.458 \\

Employment     & 3  & 4.7    & 0.0012  & 0.934 \\
Employment     & 6  & 14.3   & 0.0088  & 0.676 \\
Employment     & 12 & 21.5   & 0.0330  & 0.120 \\

\addlinespace

\multicolumn{5}{l}{\textit{Panel B: Monthly growth rates + month fixed effects}} \\

Output         & 3  & 0.0    & 0.0001  & 0.998 \\
Output         & 6  & -9.1   & -0.0245 & 0.058 \\
Output         & 12 & -0.3   & -0.0009 & 0.943 \\

Employment     & 3  & -0.9   & -0.0004 & 0.838 \\
Employment     & 6  & -1.1   & -0.0004 & 0.626 \\
Employment     & 12 & 2.9    & 0.0034  & 0.088 \\

\bottomrule
\end{tabular}

\begin{tablenotes}
\footnotesize
\item Notes: The table reports out-of-sample forecasting performance from direct forecasting regressions. ``RMSE gain'' denotes the percentage reduction in root mean squared forecast error from augmenting an autoregressive benchmark with incorporation growth. Positive values indicate improved forecast performance. HAC standard errors use Newey--West corrections with lag length equal to the forecast horizon.
\end{tablenotes}

\end{threeparttable}
\end{table}

Table \ref{tab:forecast_results} shows that incorporation growth improves forecasts of employment growth, particularly at the 12-month horizon where RMSE falls by 21.5\%. Forecasting gains are considerably weaker for output and largely disappear when month-on-month growth rates are used. Consistent with the lead--lag results, the predictive content of incorporations therefore appears concentrated in lower-frequency cyclical dynamics rather than high-frequency fluctuations. One possible explanation for the weaker output results is that monthly GVA is measured with greater short-run noise than PAYE-based employment.

Importantly, the predictive relationship between incorporation activity and future employment does not require that the incorporated firms themselves subsequently account for the observed employment growth, although our evidence that incorporations lead official business births is consistent with this mechanism. A broader interpretation is that both incorporation activity and future employment respond to common expectations about future economic conditions. For example, improving demand conditions may lead entrepreneurs to incorporate new businesses immediately, while incumbent firms expand employment only gradually over subsequent quarters. Under this interpretation, incorporation activity serves as an early real-time signal of evolving economic conditions, regardless of whether the newly incorporated firms themselves become major employers.

\section{Dynamic Effects of Firm Entry}
\label{sec:var}

We study the dynamic effects of firm entry shocks on employment and output using Cholesky-identified SVAR. The objective is to quantify how employment and output respond following shocks to firm creation.

We estimate two complementary VAR specifications using either year-on-year growth rates,
$\mathbf{y}^{y}_t = 100[\Delta_{12}\log E_t,\Delta_{12}\log L_t,\Delta_{12}\log Y_t]'$,
or month-on-month growth rates,
$\mathbf{y}^{m}_t = 100[\Delta\log E_t,\Delta\log L_t,\Delta\log Y_t]'$.
Month fixed effects are included in the month-on-month specification to control for seasonality.
\[
\mathbf{y}_t
=
\sum_{i=1}^{p} A_i \mathbf{y}_{t-i}
+
\mathbf{u}_t,
\]
with the lag length fixed at $p=6$.

To identify dynamic responses, we use a recursive (Cholesky) ordering $E \rightarrow L \rightarrow Y$. This ordering reflects the forward-looking nature of firm formation and institutional lags in the registration process, which make entry unlikely to respond contemporaneously to macroeconomic conditions within the month. Employment is allowed to respond contemporaneously to entry shocks, while output may respond contemporaneously to both.

The two specifications capture different frequencies of variation. The year-on-year specification smooths high-frequency volatility and highlights persistent medium-run movements in entry and macroeconomic activity. The month-on-month specification instead focuses on higher-frequency fluctuations and permits cumulative responses to be interpreted approximately as level effects. Our discussion therefore focuses primarily on the year-on-year specification, with the month-on-month results providing a complementary perspective on shorter-run dynamics.

\subsection{Year-on-Year Impulse Responses}
\label{sec:var_results}

\begin{figure}[H]
\centering
\includegraphics[width=0.95\textwidth]{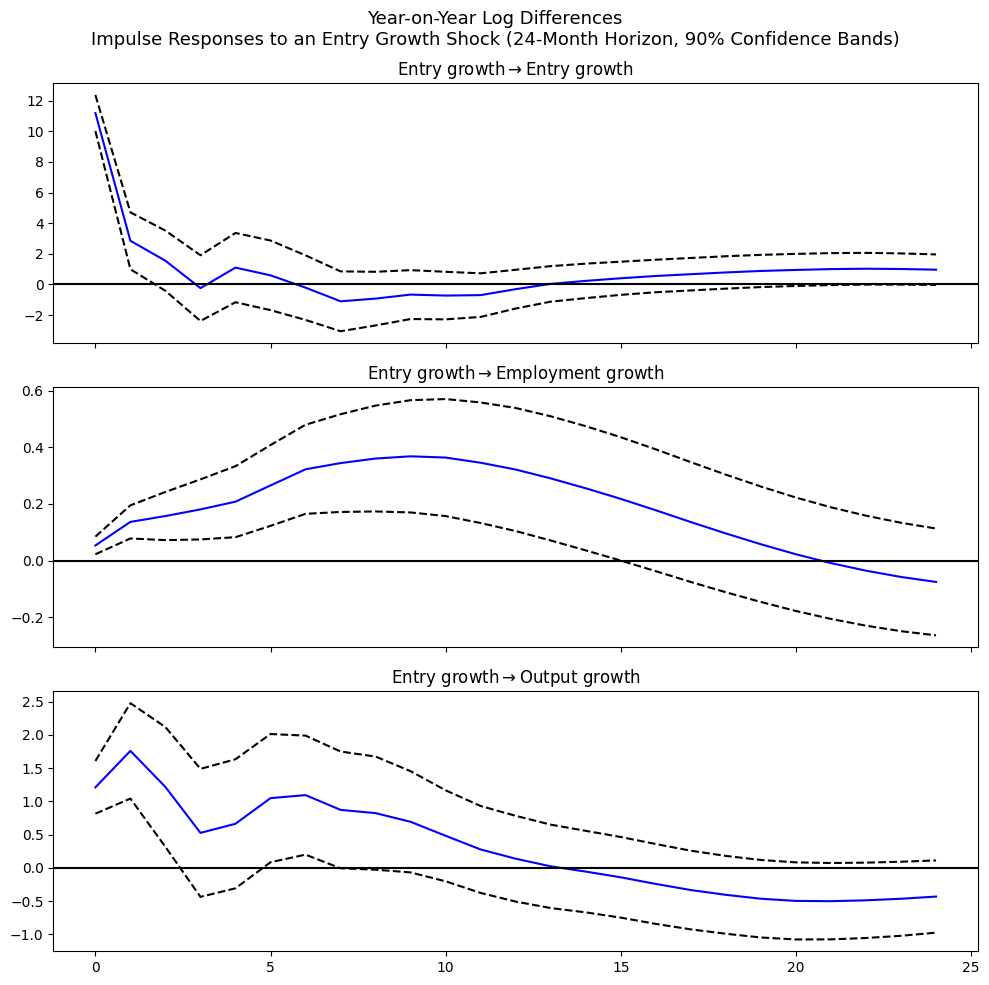}
\caption{Impulse responses to an entry growth shock: year-on-year log differences}
\label{fig:irf_yoy_main}
\end{figure}

Figure \ref{fig:irf_yoy_main} reports impulse responses from the benchmark VAR estimated using year-on-year log differences. A positive shock to firm entry growth generates a persistent increase in employment growth. The response rises gradually, peaks after approximately 8--12 months, and then slowly dissipates. Quantitatively, a one-standard-deviation shock to year-on-year entry growth raises annual employment growth by approximately 0.4\% after twelve months. These dynamics are consistent with a delayed transmission mechanism in which newly incorporated firms gradually become economically active through hiring and expansion.

Output growth also increases following an entry shock, with positive responses emerging at medium horizons. Although the output response is estimated less precisely than the employment response, the magnitudes remain economically meaningful and are consistent with sustained waves of firm creation contributing to broader macroeconomic activity over time.

\subsection{Month-on-Month Impulse Responses}
\label{sec:mom_var_results}

The month-on-month specification focuses on higher-frequency fluctuations in firm creation. Because the VAR is estimated in monthly log differences, we present cumulative impulse responses, which approximately recover the implied effects on employment and output levels. We present month-on-month levels responses in Appendix~\ref{app:irf_interpretation}.

\begin{figure}[H]
\centering
\includegraphics[width=0.95\textwidth]{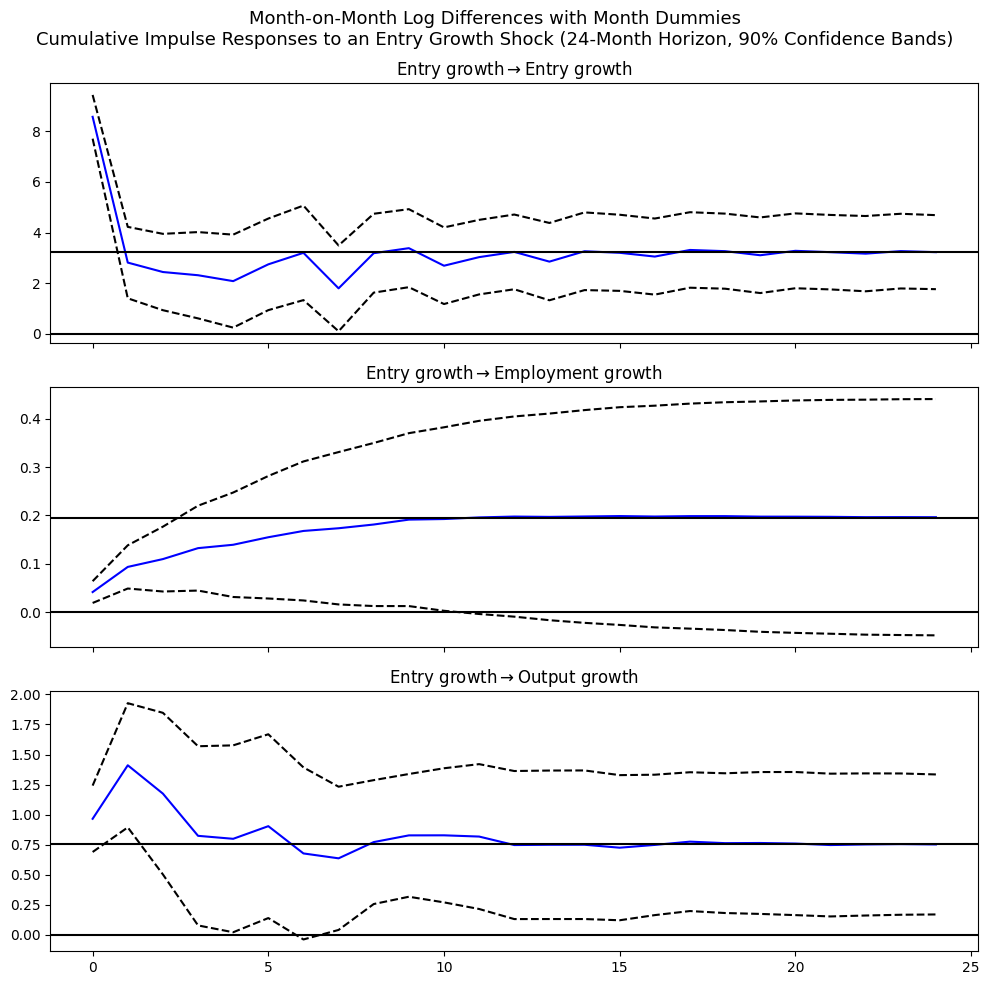}
\caption{Cumulative impulse responses to an entry growth shock: month-on-month log differences}
\label{fig:irf_mom_cumulative}
\end{figure}
Comparing Figures \ref{fig:irf_yoy_main} and \ref{fig:irf_mom_cumulative} highlights the distinction between temporary fluctuations and more persistent changes in entrepreneurial activity. While the month-on-month specification implies a cumulative employment increase of around 0.2\% after twelve months, Figure \ref{fig:irf_yoy_main} shows that a one-standard-deviation shock to year-on-year entry growth raises annual employment growth by approximately 0.4\% after one year. This difference is intuitive because the year-on-year specification captures a broader and more persistent increase in firm creation than a one-off month-on-month surge. The larger response therefore suggests that sustained waves of firm formation contain additional information about underlying macroeconomic conditions and transmit more strongly through hiring and production decisions.

\section{Conclusion}
\label{sec:conclusion}
This paper introduces Companies House Real-Time (CHRT), a novel administrative dataset that records firm creation and dissolution activity for the full population of UK-incorporated companies at high frequency. We show that CHRT entry data closely track official business demography statistics while becoming available substantially earlier, providing a timely measure of business formation.

Using monthly and quarterly data, we document the behaviour of firm entry and estimate the macroeconomic effects of entry shocks. We find that increases in firm entry are followed by persistent and statistically significant gains in employment and output.

The predictive value of incorporation activity does not depend on every newly incorporated company becoming an active employer business. Rather, firm creation appears to contain forward-looking information about future economic conditions, whether because some new firms subsequently generate employment and output, because entrepreneurs respond to information about future prospects, or both. More broadly, our results suggest that real-time firm entry data can complement existing official statistics and improve the monitoring of macroeconomic conditions.

\newpage
\printbibliography

\newpage
\appendix

\counterwithin{figure}{section}
\counterwithin{table}{section}
\counterwithin{equation}{section}

\section{Companies House Real-Time Dataset}
\label{sec:appendix_chrt}

We construct a longitudinal census of incorporated companies in the United Kingdom using Companies House (CH) records. The resulting Companies House Real-Time (CHRT) dataset combines historical register snapshots with Companies House API data to track firm entry, exit, industrial classification, and geographic location over time.

\subsection{Data Sources}

The primary source is the Companies House ``Free Company Data'' register, archived through the UK Government Web Archive since July 2012. Historical snapshots are intermittent prior to 2020 and become approximately monthly thereafter. We complement these snapshots with Companies House API data containing incorporation and dissolution dates.

For each company, we extract the Company Registration Number (CRN), incorporation date, dissolution date where applicable, postcode, and five-digit SIC industry classifications.

\subsection{Panel Construction}

We construct a longitudinal panel indexed by company and register release date. Firm entries are identified as incorporations appearing between successive register snapshots, while firm exits are dated using official dissolution records from the Companies House API. Repeated snapshots additionally allow us to track changes in firms' SIC classifications and registered office postcodes over time.

Historical snapshots were retrieved from the UK Government Web Archive. Earlier archives required partial manual reconstruction because of changes in Companies House file naming conventions over time.

The final dataset contains more than 10 million incorporated companies and over 300 million company--snapshot observations between 2012 and 2024.

\subsection{Validation}

\begin{figure}[H]
    \centering

    \begin{subfigure}[b]{\textwidth}
        \centering
        \includegraphics[width=\textwidth]{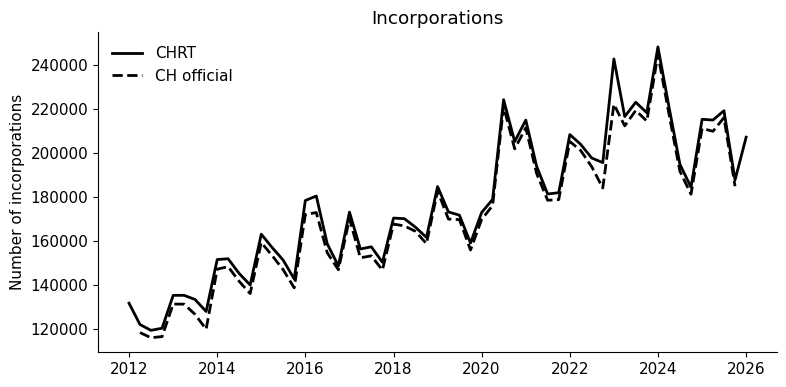}
        \caption{CH Official vs CHRT Incorporations}
        \label{fig:UK_v_UKBD_incs}
    \end{subfigure}

    \vspace{1em}

    \begin{subfigure}[b]{\textwidth}
        \centering
        \includegraphics[width=\textwidth]{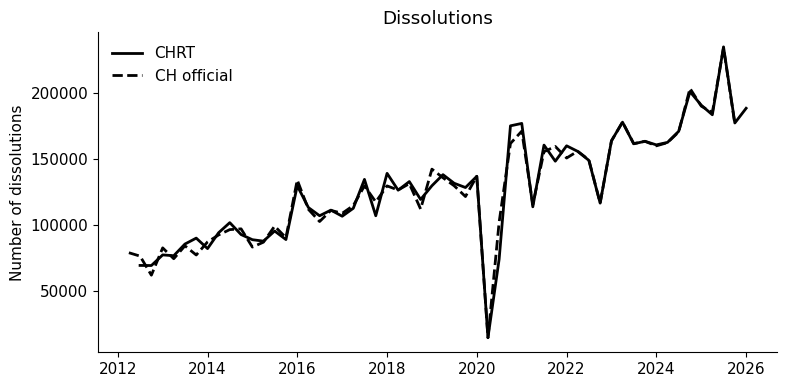}
        \caption{CH Official vs CHRT Dissolutions}
        \label{fig:UK_v_UKBD_diss}
    \end{subfigure}

    \caption{Companies House official statistics and reconstructed CHRT series}
    \label{fig:UK_v_UKBD}

    \caption*{\scriptsize \textbf{Notes:} The figures compare Companies House official statistics with the reconstructed Companies House Real-Time (CHRT) series. The close correspondence validates the reconstruction procedure. Minor differences reflect timing discrepancies, company restorations, and differences in the treatment of dissolution dates.}
\end{figure}

Figure \ref{fig:UK_v_UKBD} compares the reconstructed CHRT series against official Companies House statistics for incorporations and dissolutions. The two series closely track one another over time, supporting the accuracy of the reconstruction procedure.

\section{Additional Descriptive Evidence}
\label{app:descriptive}
Tables \ref{tab:incl_covid_stats} and \ref{tab:desc_stats_nocovid} report descriptive statistics for the full sample and for a sample excluding the COVID-19 period (March 2020--June 2021), respectively.

\begin{table}[H]
\centering
\caption{Descriptive statistics}
\label{tab:incl_covid_stats}
\begin{tabular}{lrrrrrrrr}
\toprule
Variable & Mean & Std. Dev. & P10 & Median & P90 & Min & Max & N \\
\midrule
$E_t$                     & 62,094   & 10,894   & 48,822   & 60,125   & 77,900   & 40,254   & 98,658   & 140 \\
$L_t$                     & 29.0m    & 1.1m     & 27.4m    & 29.0m    & 30.3m    & 26.8m    & 30.5m    & 140 \\
$Y_t$                     & 95.6     & 6.5      & 89.0     & 96.0     & 101.2    & 72.7     & 103.0    & 140 \\
\midrule
$100 \times \Delta \log E_t$      & 0.1      & 15.0     & -17.7    & 0.3      & 18.2     & -48.6    & 37.1     & 139 \\
$100 \times \Delta \log L_t$      & 0.1      & 0.7      & -0.4     & 0.1      & 0.5      & -1.6     & 1.0      & 139 \\
$100 \times \Delta \log Y_t$      & 0.1      & 3.0      & -1.0     & 0.2      & 1.0      & -23.3    & 11.0     & 139 \\
\midrule
$100 \times \Delta_{12}\log E_t$  & 3.0      & 14.5     & -13.0    & 2.9      & 18.7     & -36.9    & 47.6     & 128 \\
$100 \times \Delta_{12}\log L_t$  & 1.0      & 2.3      & -0.8     & 1.3      & 3.2      & -3.6     & 5.6      & 128 \\
$100 \times \Delta_{12}\log Y_t$  & 2.1      & 5.8      & -2.0     & 2.3      & 6.2      & -20.8    & 20.0     & 128 \\
\bottomrule
\end{tabular}

\vspace{0.3em}
\footnotesize
\textit{Notes:} $E_t$ denotes monthly firm entry, $L_t$ PAYE employment, and $Y_t$ monthly GVA. Growth rates are expressed in log differences and multiplied by 100, so values are approximately percentage changes. Month-on-month growth is defined as $\Delta \log X_t = \log X_t - \log X_{t-1}$, while year-on-year growth is defined as $\Delta_{12}\log X_t = \log X_t - \log X_{t-12}$.
\end{table}

\begin{table}[H]
\centering
\caption{Descriptive statistics excluding the COVID-19 period}
\label{tab:desc_stats_nocovid}
\begin{tabular}{lrrrrrrrr}
\toprule
Variable & Mean & Std. Dev. & P10 & Median & P90 & Min & Max & N \\
\midrule
$E_t$                     & 61,015   & 10,785   & 48,888   & 58,985   & 76,344   & 40,254   & 98,658   & 124 \\
$L_t$                     & 29.0m    & 1.1m     & 27.5m    & 28.9m    & 30.4m    & 26.8m    & 30.5m    & 124 \\
$Y_t$                     & 96.4     & 4.5      & 89.2     & 97.1     & 101.8    & 87.9     & 103.0    & 124 \\
\midrule
$100 \times \Delta \log E_t$      & 0.1      & 14.9     & -17.2    & -0.3     & 16.8     & -48.6    & 37.1     & 123 \\
$100 \times \Delta \log L_t$      & 0.1      & 0.1      & -0.0     & 0.1      & 0.3      & -0.1     & 0.6      & 123 \\
$100 \times \Delta \log Y_t$      & 0.1      & 0.3      & -0.3     & 0.1      & 0.5      & -0.7     & 1.0      & 123 \\
\midrule
$100 \times \Delta_{12}\log E_t$  & 1.3      & 11.9     & -14.5    & 3.1      & 14.8     & -36.9    & 37.7     & 112 \\
$100 \times \Delta_{12}\log L_t$  & 1.5      & 1.1      & -0.0     & 1.3      & 2.8      & -0.4     & 4.5      & 112 \\
$100 \times \Delta_{12}\log Y_t$  & 2.2      & 2.3      & 0.6      & 1.5      & 3.7      & -0.5     & 11.9     & 112 \\
\bottomrule
\end{tabular}

\vspace{0.3em}
\footnotesize
\textit{Notes:} The COVID-19 period (March 2020 to June 2021) is excluded from the sample. $E_t$ denotes monthly firm entry, $L_t$ PAYE employment, and $Y_t$ monthly GVA. Growth rates are expressed in log differences and multiplied by 100, so values are approximately percentage changes.
\end{table}

\subsection{Time-Series Plots}
\label{app:timeseries}

\begin{figure}[H]
\centering
\includegraphics[width=\textwidth]{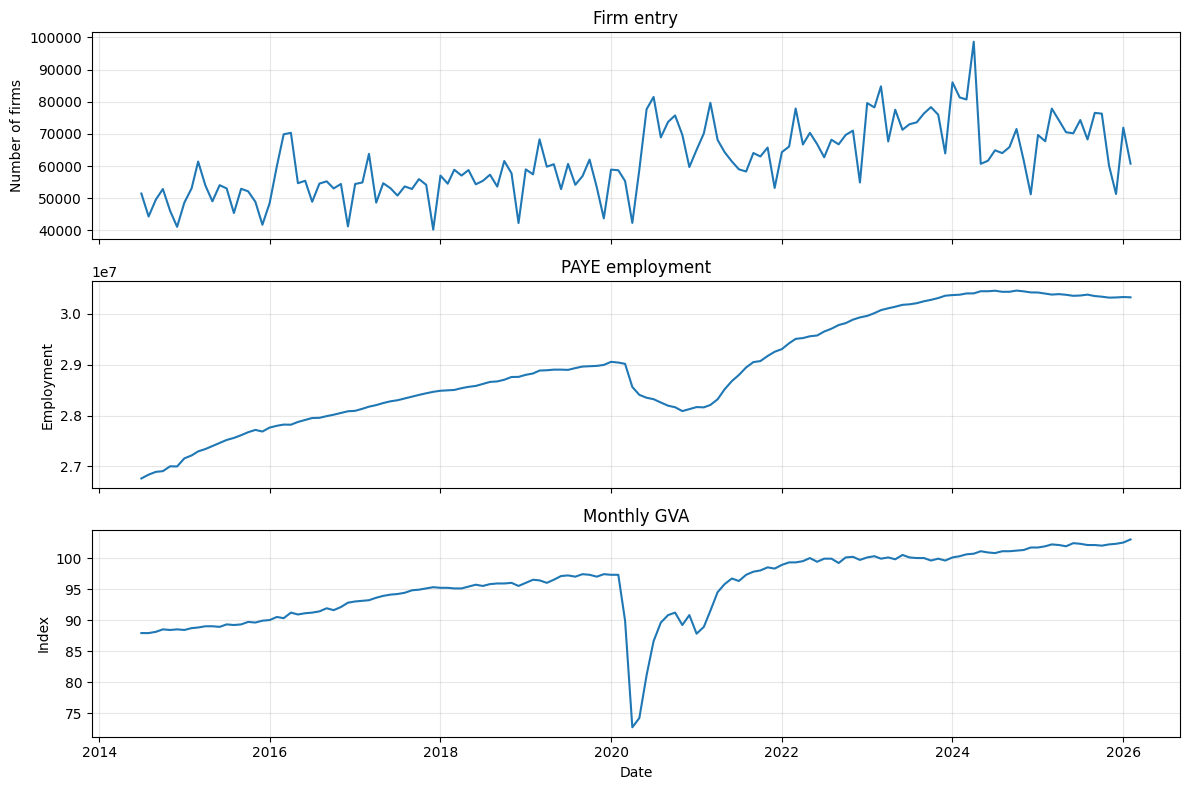}
\caption{Levels of firm entry, employment, and output}
\label{fig:levels}
\end{figure}

\begin{figure}[H]
\centering
\includegraphics[width=\textwidth]{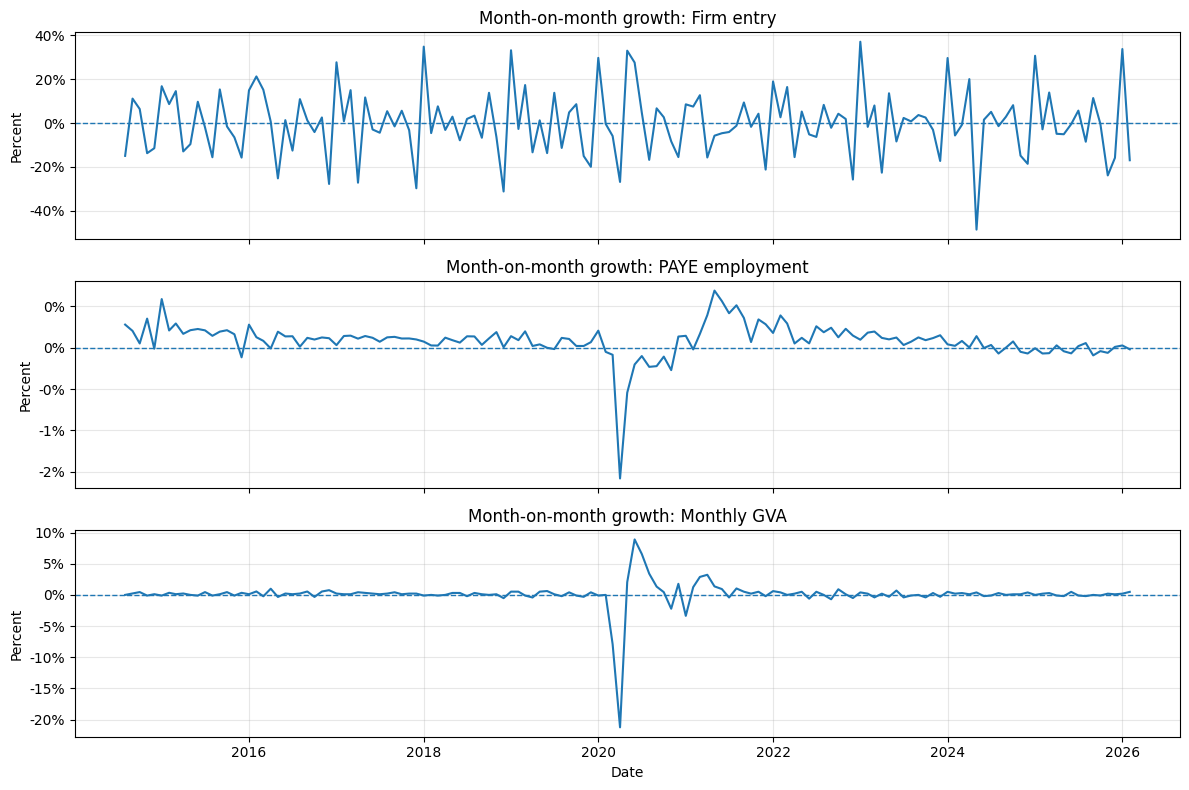}
\caption{Month-on-month growth rates}
\label{fig:mom_growth}
\end{figure}

\begin{figure}[H]
\centering
\includegraphics[width=\textwidth]{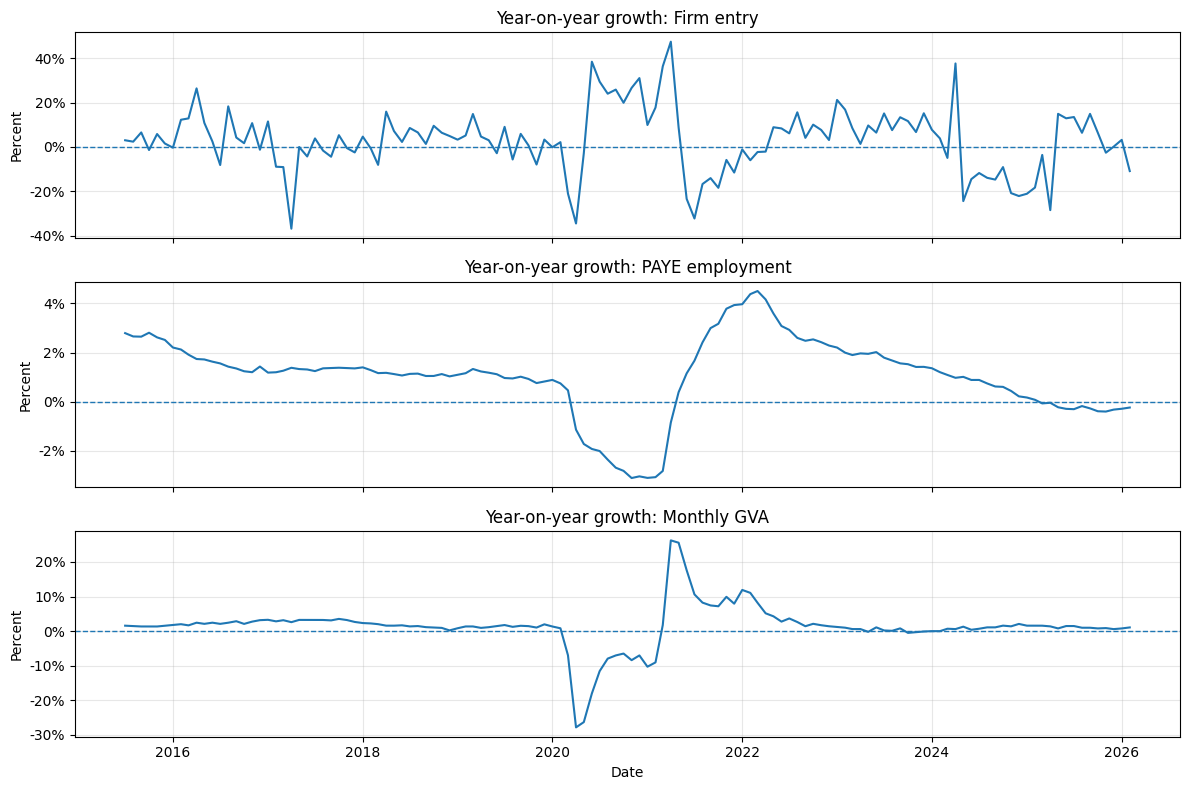}
\caption{Year-on-year growth rates}
\label{fig:yoy_growth}
\end{figure}

Figures \ref{fig:levels}--\ref{fig:yoy_growth} plot firm entry, employment, and output in levels, month-on-month growth rates, and year-on-year growth rates.

\section{Additional Impulse Response Specifications}
\label{app:irf_interpretation}

\begin{figure}[H]
\centering
\includegraphics[width=0.95\textwidth]{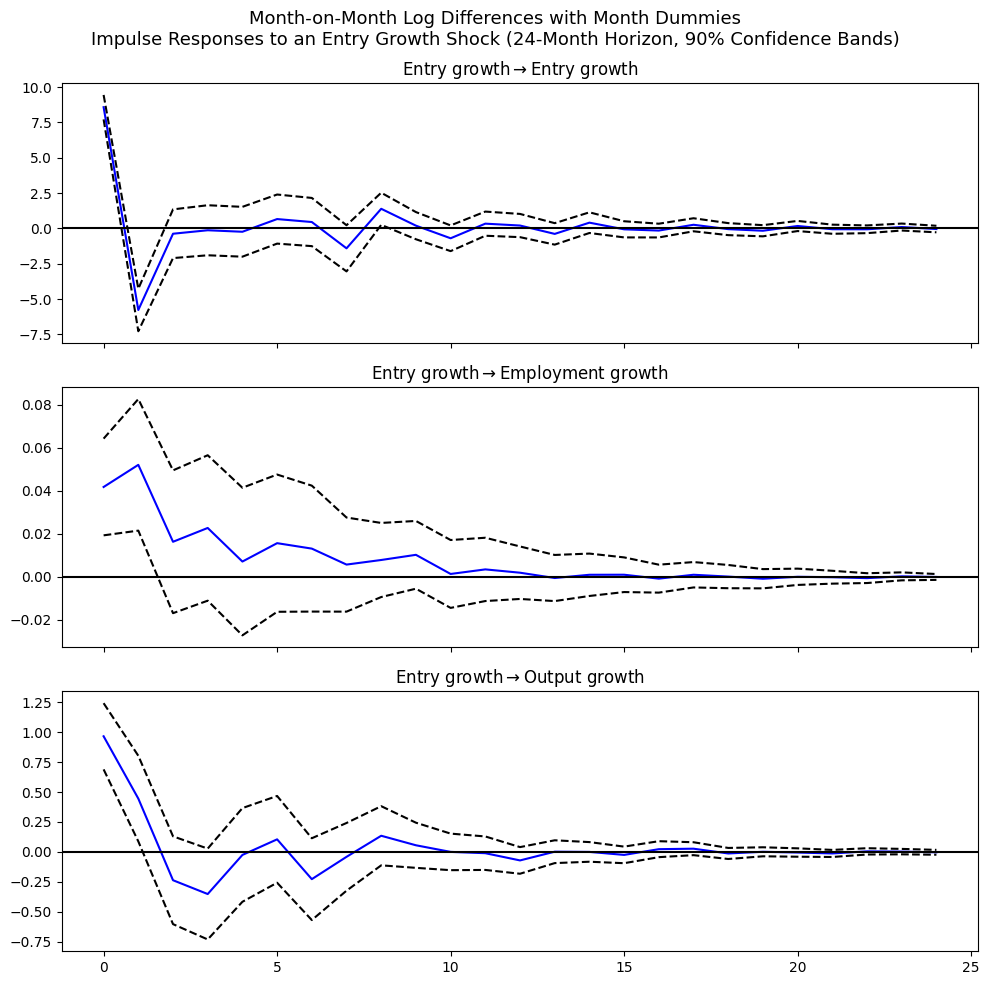}
\caption{Impulse responses to an entry growth shock: month-on-month log differences}
\label{fig:irf_mom_level}
\end{figure}

Figure \ref{fig:irf_mom_level} reports impulse responses from the VAR estimated in month-on-month log differences. In this specification, the horizon-specific impulse responses measure the effect of an entry shock on monthly employment and output growth rates.

\end{document}